# The COMICS Tool – <u>C</u>omputing <u>M</u>inimal <u>C</u>ounterexample<u>s</u> for DTMCs[*]

Nils Jansen[1], Erika Ábrahám[1], Maik Scheffler[1], Matthias Volk[1], Andreas Vorpahl[1], Ralf Wimmer[2], Joost-Pieter Katoen[1], and Bernd Becker[2]

[1] RWTH Aachen University, Germany
[2] Albert-Ludwigs-University Freiburg, Germany

**Abstract.** This report presents the tool `COMICS` 1.0, which performs model checking and generates counterexamples for DTMCs. For an input DTMC, `COMICS` computes an abstract system that carries the model checking information and uses this result to compute a *critical subsystem*, which induces a counterexample. This abstract subsystem can be refined and concretized *hierarchically*. The tool comes with a command-line version as well as a graphical user interface that allows the user to interactively influence the refinement process of the counterexample.

## 1 Introduction

*Discrete-time Markov chains* (DTMCs) are widely used to model safety-critical systems with uncertainties. Model checking *probabilistic computation tree logic* (PCTL) [1] properties can be performed by prominent tools like PRISM [2] and MRMC [3]. Unfortunately, the implemented numerical methods do not provide diagnostic information in form of *counterexamples*, which are very important for debugging and are also needed for CEGAR frameworks [4].

Although different approaches were proposed for probabilistic counterexamples ([5,6,7,8,9]), there is still a lack of efficient and user-friendly *tools*. To fill this gap, we developed the tool `COMICS` (<u>Co</u>mputing <u>M</u>inimal <u>C</u>ounterexample<u>s</u>), supporting SCC-based model checking [10] and, in case the property is violated, the *automatic* generation of *abstract counterexamples* [8], which can subsequently be refined either automatically or guided by the user.

While most approaches represent probabilistic counterexamples as sets of paths, we use (hierarchically abstracted) subgraphs of the input DTMC, so-called *critical subsystems*. This allows for a much more *compact representation* and a *significant decrease in the computational complexity*. The user can refine abstract critical subsystems *hierarchically* by choosing system parts of interest which are to be concretized and further examined. All computation steps of the hierarchical counterexample refinement can be *guided and revised*. Though refinement can be done until a fully concrete counterexample is gained, it seems likely that the user can gain sufficient debugging information from abstract systems considering real-world examples with millions of states. The tool comes with a graphical user interface (GUI) which permits the *visualization and reviewing* of existing test cases and the *creation* of random examples as well as new test cases.

---

[*] This work was partly supported by the German Research Council (DFG) as part of the research project CEBug (AB 461/1-1), the Transregional Collaborative Research Center AVACS (SFB/TR 14) and the Research Training Group AlgoSyn (1298) as well as by the Netherlands Organisation for Scientific Research (NWO) as part of the DFG/NWO Bilateral Research Programme ROCKS.

The implemented methods result in substantial improvements regarding the size and the number of computation steps for the generation of probabilistic counterexamples. The only other available tool we are aware of is DiPro [11]. However, it does not support abstract counterexamples, which is crucial for the handling of large systems. It also does not allow the user to influence the search process by using his or her expertise. Comparative experiments show that we can compute reasonably smaller counterexamples in shorter time with our tool.

In Section 2 we recall some preliminaries regarding DTMCs and counterexamples. In Section 3 we give a brief introduction to the methods implemented in our tool. We describe the features and architecture in Section 4 and report on some benchmarks . We conclude the paper in Section 5. The tool, a detailed manual, and a number of benchmarks are available at the COMICS website[3].

## 2 Preliminaries

In this section we give some basic foundations. See [12] for more details.

**Definition 1.** *Assume a set AP of atomic propositions. A* discrete-time Markov chain (DTMC) *is a tuple* $M = (S, I, P, L)$ *with a non-empty finite state set $S$, an initial discrete probability distribution $I : S \to [0, 1]$ with $\sum_{s \in S} I(s) = 1$, a transition probability matrix $P : S \times S \to [0, 1]$ with $\sum_{s' \in S} P(s, s') = 1$ for all $s \in S$, and a labeling function $L : S \to 2^{AP}$.*

Assume in the following a set $AP$ of atomic propositions and a DTMC $M = (S, I, P, L)$.

We say that there is a *transition* from a state $s \in S$ to a state $s' \in S$ iff $P(s, s') > 0$. A *path* of $M$ is a finite or infinite sequence $\pi = s_0 s_1 \ldots$ of states $s_i \in S$ such that $P(s_i, s_{i+1}) > 0$ for all $i$. We say that the transitions $(s_i, s_{i+1})$ are *contained* in the path $\pi$, written $(s_i, s_{i+1}) \in \pi$. We write $Paths_{inf}^M$ for the set of all infinite paths of $M$, and $Paths_{inf}^M(s)$ for those starting in $s \in S$. Analogously, $Paths_{fin}^M$ is the set of all finite paths of $M$, $Paths_{fin}^M(s)$ of those starting in $s$, and $Paths_{fin}^M(s, t)$ of those starting in $s$ and ending in $t$. A state $t$ is called *reachable* from another state $s$ iff $Paths_{fin}^M(s, t) \neq \emptyset$.

A state set $S' \subseteq S$ is called *absorbing in $M$* iff there is a state in $S'$ from which no state outside $S'$ is reachable in $M$. We call $S'$ *bottom in $M$* if this holds for all states in $S'$. States $s \in S$ with $P(s, s) = 1$ are also called *absorbing states*.

---

[3] http://www-i2.informatik.rwth-aachen.de/i2/comics/

We call $M$ *loop-free*, if all of its loops are self-loops on absorbing states. A set $S' \subseteq S$ is *strongly connected in $M$* iff for all $s, t \in S'$ there is a path from $s$ to $t$ visiting states from $S'$ only. A *strongly connected component (SCC)* of $M$ is a maximal strongly connected subset of $S$.

The probability measure for finite paths $\pi \in Paths_{fin}^M$ is defined by $Pr_{fin}^M(\pi) = \prod_{(s_i, s_{i+1}) \in \pi} P(s_i, s_{i+1})$. For a set $R \subseteq Paths_{fin}^M$ of paths we have $Pr_{fin}^M(R) = \sum_{\pi \in R'} Pr_{fin}^M(\pi)$ with $R' = \{\pi \in R \mid \forall \pi' \in R.\ \pi'$ is no prefix of $\pi\}$.

The syntax of *probabilistic computation tree logic (PCTL)* [13] is given by[4]

$$\varphi ::= p \mid \neg\varphi \mid \varphi \wedge \varphi \mid \mathbb{P}_{\sim\lambda}(\varphi\ U\ \varphi)$$

for (state) formulae with $p \in AP$, $\lambda \in [0, 1] \subseteq \mathbb{R}$, and $\sim\ \in \{<, \leq, \geq, >\}$. We define the "finally"-operator ($\Diamond$) and the "globally"-operator $\square$ in the usual way.

For a property $\mathbb{P}_{\leq\lambda}(\varphi_1\ U\ \varphi_2)$ refuted by $M$, a *counterexample* is a set $C \subseteq Paths_{fin}^M$, $Pr_{fin}^M(C) > \lambda$ of finite paths starting in an initial state and *satisfying* $\varphi_1\ U\ \varphi_2$. For $\mathbb{P}_{<\lambda}(\varphi_1\ U\ \varphi_2)$, the probability mass has to be at least $\lambda$. We consider upper probability bounds only; see [6] for the reduction of lower bounds to this case.

We reduce the problem of checking probabilistic until properties $\mathbb{P}_{\leq\lambda}(\varphi_1\ U\ \varphi_2)$ to probabilistic reachability problems as follows: Each state of the DTMC $M$ that satisfies $(\neg\varphi_1 \wedge \neg\varphi_2) \vee \varphi_2$ is made absorbing. Instead of checking $\mathbb{P}_{\leq\lambda}(\varphi_1\ U\ \varphi_2)$ for $M$, we check $\mathbb{P}_{\leq\lambda} \Diamond \varphi_2$ for the modified DTMC. The $\varphi_2$-states are also called *target* states. We concentrate on this reduced problem and assume DTMCs to have single initial and target states. Note, that each DTMC can be equivalently transformed w. r. t. an $U$-formula to satisfy these requirements.

## 3 Hierarchical Counterexamples

In [10] we proposed a model checking approach for DTMCs based on *hierarchical abstraction*. Each SCC of the underlying graph of the input DTMC is abstracted by a state whose outgoing transitions lead to states outside the SCC and carry the whole probability mass of reaching those states when once entering the SCC. This abstraction is done recursively in a bottom-up manner: before abstracting an SCC we first apply abstraction to the sub-SCCs nested in it. The final result is an abstract DTMC whose only transitions lead from the initial state of the input DTMC to absorbing states and carry the corresponding reachability probabilities.

Fig. 1(a) depicts a DTMC and its nested SCC structure: SCC $S_1$ contains SCC $S_{1.1}$. The upper graph of Fig. 1(b) depicts the result of the model checking: The probability to reach the target state 3 from the initial state is 0.9. This hierarchically abstracted DTMC can be also hierarchically concretized. The lower graph of Fig. 1(b) shows the concretization of the abstract state $s_0$: The outgoing edges of $s_1$ carry the probability mass of all paths leading from the input state 4 of the SCC $S_1$ to the output states 3 and 7, respectively. Fig. 1(c) shows a

---

[4] In this paper we only consider unbounded properties.

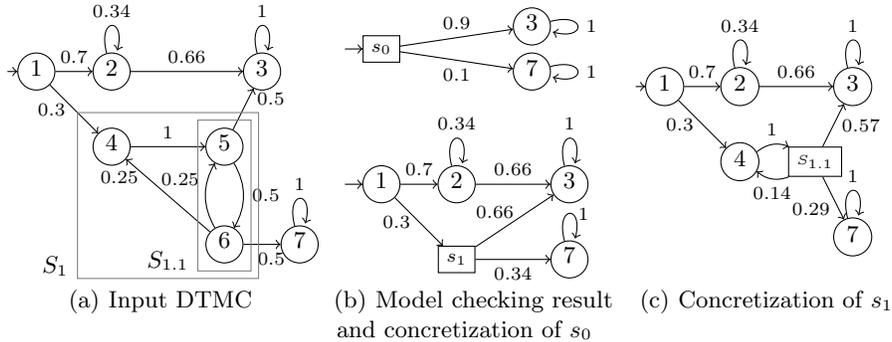

(a) Input DTMC  (b) Model checking result  (c) Concretization of $s_1$
              and concretization of $s_0$

**Fig. 1.** Example SCC-based model checking

further concretization step: the state $s_1$ is concretized while the SCC $S_{1.1}$ is still abstracted. Concretizing also the abstract state $s_{1.1}$ would result in the DTMC of Figure 1(a). Based on this approach, we presented in [8] a method to compute and represent counterexamples as *critical subsystems*, consisting of subsets of the original DTMC's states and transitions such that the probability of reaching target states from the initial state within the subsystem still exceeds the probability bound $\lambda$. The method first computes an abstract critical subsystem for the abstract DTMC resulting from model checking. Inside this abstract DTMC one or more abstract states are selected and concretized, and a critical subsystem is determined for the concretized system. This process may be repeated until the system is fully concretized. We suggested two methods for the computation of critical subsystems: The *global search* (`GS`) looks for most probable paths through the whole system until the involved states and transitions form a critical subsystem. The *local search* (`LS`) builds critical subsystems incrementally by extending subsystems with most probable path fragments.

Its application to benchmarks showed the competitiveness of the SCC-based model checking. Compared to other approaches, experiments for the counterexample generation revealed an improvement by several orders of magnitude in the number of paths needed to form the counterexample as well as in the number of involved states.

## 4 The COMICS Tool

`COMICS` can be used either as a command-line tool or with a GUI, the latter allowing the user to actively influence the process of finding a counterexample. We therefore distinguish between *command-line mode* and *interactive mode*. The program consists of approximately 20 000 lines of code. The GUI is implemented in `Java`, all other components in `C++`. The user may select *exact* or *floating point* arithmetics for the computations.

In command-line mode, *SCC-based model checking* can be performed for an input DTMC and a PCTL property. If model checking reveals that the probability bound is exceeded, a counterexample can either be computed on the *abstract*

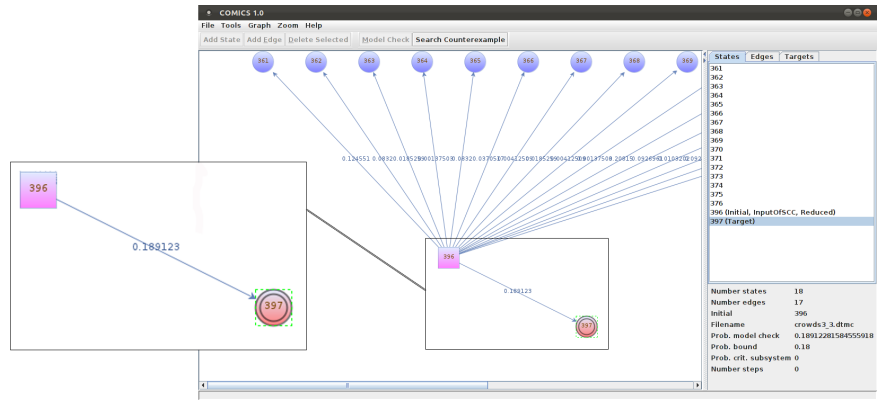

**Fig. 2.** Screenshot of COMICS's GUI with an instance of the crowds protocol

*system and refined hierarchically* or a counterexample can directly be computed on the *concrete system*. In the first case, *heuristics* for the number of states to be concretize in a single step as well as for the choice of states are offered. It is also possible to predefine the number of concretization steps. The user can choose between the counterexample representation as a *set of paths* and as a *critical subsystem*. In the first case, the tool uses the global search and computes a *minimal counterexample* as introduced in [6]. In the second case, both the global search and the local search can be applied. For measuring the performance of the particular functions, several predefined *benchmarking options* are provided.

The interactive mode is based on the usage of the GUI. It provides a *graph editor* for specifying and modifying DTMCs. Several *layout algorithms* increase the usability even for large graphs. Both concrete and abstract graphs can be *stored*, *loaded*, *abstracted*, and *concretized* by the user. After calling SCC-based model checking, the resulting refinable abstract graph is visualized and the *counterexample generation* is invokable. As most important feature, the user is able to *control the hierarchical concretization* of a counterexample. Abstract states can either be *concretized by the user* or *automatically by heuristics*. If an input graph seems to be too large to display, the tool offers to operate without the graphical representation.

Figure 2 shows one abstracted instance of the *crowds protocol* benchmark [14], where the probability of reaching the unique target state is displayed in the information panel on the right as well as on the edge leading from the initial state to the target state. The initial state is abstract and can therefore be expanded.

The tool's five core software components are depicted in Figure 3. The functionalities and interactions are as follows:

SccMC performs SCC-based model checking for an input DTMC. An abstract DTMC is returned either to Concretize or to GUI.
Concretize decides based on either user preferences or heuristics if some nodes are to be concretized. The possibly modified system is returned to the GUI component for further user input or to CritSubSys.

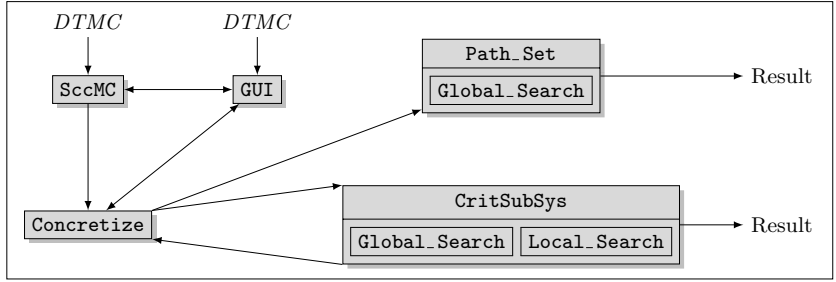

**Fig. 3.** Architecture of COMICS

- CritSubSys computes a critical subsystem using the global search or the local search. The resulting subsystem is given to Concretize for further refinement or returned as result.
- Path_Set uses the global search approach to compute a set of paths which forms a minimal counterexample.
- GUI provides the possibility to define DTMCs by their underlying graphs and to modify them using the JGraph library[15]. This component is connected both to SccMC and Concretize via Java Native Interface (JNI).

The original explicit input format for DTMCs is adapted from Mrmc. For the abstract graphs we defined an XML-format which efficiently stores their tree-like hierarchical structure. Using XML-parsers, this allows for transferring the graph data between the different components and save it for later processing. An import of Mrmc input files is provided. Thus the export facilities of Prism allow us to test a large number of benchmarks offered on the Prism-website [16].

|       |           | crowds |  |  |  |  |  | contract signing |  |  |  |
|-------|-----------|--------|--------|--------|--------|--------|--------|--------|--------|--------|--------|
| states |          | 3515   | 3515   | 18817  | 198199 | 485941 | 1058353 | 33790 | 156670 | 737278 | 1654782 |
| transitions |      | 6035   | 6035   | 32677  | 198199 | 857221 | 1872313 | 34813 | 157693 | 753663 | 1671165 |
| total prob. |      | 0.2346 | 0.2346 | 0.4270 | 0.7173 | 0.809  | 0.8731 | 0.5156 | 0.5156 | 0.5039 | 0.5039 |
| prob. threshold |  | 0.15   | 0.23   | 0.25   | 0.25   | 0.35   | 0.4    | 0.4    | 0.5    | 0.5    | 0.5    | 0.5 |
| GS | # states | 629 | 1071 | 2036 | 5198 | 5248 | 5250 | 6827 | 37601 | 140034 | 369448 |
|    | prob.    | 0.1501 | 0.2301 | 0.25 | 0.3503 | 0.4002 | 0.4001 | 0.5 | 0.5 | 0.5 | 0.5 |
|    | time (s) | 0.02 | 0.38 | 0.38 | 7.97 | 16.36 | 18.78 | 0.36 | 2.98 | 238.82 | 605.81 |
| LS | # states | 182 | 900 | 943 | 4180 | 6368 | – | 6657 | 37377 | MO | MO |
|    | prob.    | 0.1501 | 0.2302 | 0.2501 | 0.3501 | 0.4 | – | 0.5 | 0.5 | – | – |
|    | time (s) | 0.14 | 1.11 | 6.1 | 619.06 | 2455.46 | TO | 8 | 54.58 | – | – |
| $k$SP | # states | 1071 | – | – | – | – | – | 6827 | 37601 | 140034 | 369444 |
|       | prob.    | 0.15 | – | – | – | – | – | 0.5 | 0.5 | 0.5 | 0.5 |
|       | time (s) | 6.58 | TO | TO | TO | TO | TO | 1.93 | 0.13 | 0.69 | 1.49 |
| DiPro | # states | 938 | 2901 | 3227 | 9005 | – | – | 13311 | 74751 | MO | MO |
|       | prob.    | 0.1675 | 0.2334 | 0.254 | 0.3533 | – | – | 0.5 | 0.5 | – | – |
|       | time (s) | 2.02 | 7.06 | 7.87 | 44.34 | ERR | ERR | 1210 | 7114 | – | – |

**Fig. 4.** Results for crowds and contract signing (TO > 2h)

Fig. 4 provides a comparison with DiPro [11]. We applied our tool using GS, LS and the $k$-shortest path ($k$SP) approach [6] to the *crowds protocol* and the

*probabilistic contract signing protocol* [17]. For the models, we give the *number of states*, the *number of transitions*, the *total probability* of reaching target states, and the used *probability threshold*, which shall be exceeded by a counterexample.

We measured the size of the counterexample (# states), the probability of reaching target states (prob.) and the computation time excluding the initial model checking. TO denotes timeout, MO out of memory and ERR wrong result. On the crowds protocol, `GS` performs best, while `LS` computes in general smaller counterexamples. $k$`SP` is the fastest method for contract signing, however, the representation of the result consists of a huge number of paths instead of a small subsystem of the input DTMC.

## 5 Conclusion and Future Work

We presented version 1.0 of our tool `COMICS` which generates abstract, hierarchically refinable counterexamples for DTMCs. In the future, we want to integrate the computation of *minimal* critical subsystems [9] and the adaption of our approaches to *symbolic data structures*. Currently we are working on an *incremental version of the Dijkstra algorithm* to improve the local search and on *compositional counterexamples* to increase the usability of debugging information, since PRISM models are usually built by parallel composition.

To speed-up the model checking process, we will connect our tool to PRISM and MRMC.